# Resonant electron–phonon–electron interaction

Jiang-Tao Liu[1,2]

Department of Physics, Nanchang University, Nanchang 330031, China

Nanoscale Science and Technology Laboratory, Institute for Advanced Study, Nanchang University, Nanchang 330031, China

**Abstract**: The effect of the resonance of electron scattering energy difference and phonon energy on the electron–phonon–electron interaction (EPEI) is studied. Results show that the resonance of electron transition energy and phonon energy can enhance EPEI by a magnitude of 1 to 2. Moreover, the anisotropic S-wave electron or $d_{x^2-y^2}$ electron can enhance resonance EPEI, and the self-energy correction of the electron will weaken resonance EPEI. Particularly, the asymmetrical spin-flip scattering process in k space can reduce the effect of electronic self-energy to enhance resonance EPEI.

## 1. Introduction

Electron–phonon–electron interaction (EPEI) has always been an important research issue in the field of condensed matter physics [1-12]. Mutual attraction between electrons can be expressed by regarding lattice vibration as the medium. Mutually attractive EPEI that leads to a Cooper pair explains the traditional superconducting phenomenon[1-6]. However, traditional EPEI is weaker, and the binding energy of the Cooper pair is lower. Meanwhile, a traditional superconductor has lower transition temperature. Given that explaining the high-temperature superconducting phenomenon is difficult for the traditional EPEI theory, a stronger EPEI should be found.

In the EPEI process, an electron with a wave vector of $\mathbf{k}$, which absorbs (emits) a phonon with a wave vector of $\mathbf{q}$, transits on the $\mathbf{k}+\mathbf{q}$ ($\mathbf{k}-\mathbf{q}$) state. Meanwhile, an electron with a wave vector of $\mathbf{k'}$, which emits (absorbs) this phonon, transits on the $\mathbf{k'}-\mathbf{q}$ ($\mathbf{k'}+\mathbf{q}$) state. An attractive interaction between electrons is likely to exist if the energy difference between the electron emission phonon before and after $|\varepsilon_\mathbf{k} - \varepsilon_{\mathbf{k}+\mathbf{q}}| = \Delta\varepsilon_{\mathbf{k},\mathbf{q}}$ is less than the phonon energy $\hbar\omega_\mathbf{q}$ [1-6]. Particularly, because the EPEI coefficient is proportional to $1/[\Delta\varepsilon_{\mathbf{k},\mathbf{q}} - (\hbar\omega_\mathbf{q})^2]$ [1-6], if $\Delta\varepsilon_{\mathbf{k},\mathbf{q}} - \hbar\omega_\mathbf{q} \to 0^-$, that is, EPEI is very strong when the electron energy difference $\Delta\varepsilon_{\mathbf{k},\mathbf{q}}$ is resonant with the phonon energy $\hbar\omega_\mathbf{q}$. This paper aims to study the effects on EPEI of the resonance of electron scattering energy difference $\Delta\varepsilon_{\mathbf{k},\mathbf{q}}$ and phonon energy $\hbar\omega_\mathbf{q}$.

The paper is organized as follows. Section 2 shows the key steps in the traditional EPEI theory. Section 3 discusses resonance electron–acoustic phonon–electron and electron–optical phonon–electron interactions and compares them with non-resonance interactions. In Section 4, we study the effect of anisotropic S-wave or $d_{x^2-y^2}$ electron and electronic self-energy correction on resonant EPEI (R-EPEI). The anisotropic S-wave electron or $d_{x^2-y^2}$ electron can enhance R-EPEI and that electronic self-energy correction can weaken R-EPEI. We study the effect of asymmetric spin-flip scattering process in k space on R-EPEI to eliminate the effect of resonant electronic self-energy correction. We found that the asymmetric spin-flip scattering process in k space can reduce the effect of electronic self-energy to enhance R-EPEI. The final section shows a qualitative discussion of the R-EPEI-based superconductor.

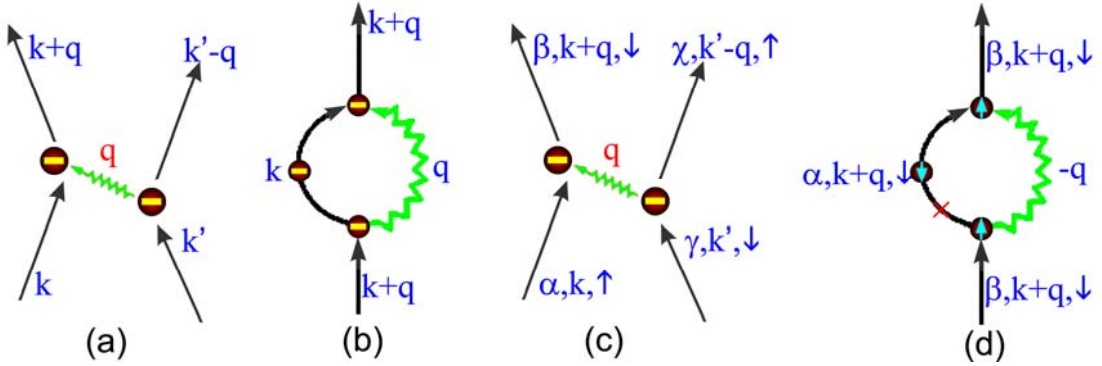

Fig. 1. Schematic diagram of the electron–phonon interaction. (a) EPEI; (b) Electron self-energy correction; (c) Spin-flip EPEI among different subbands; (d) Spin-flip electron self-energy correction among different subbands.

## 2. Theory

This section provides the key steps in traditional EPEI theory to provide a better understanding of R-EPEI. In traditional EPEI theory, the total Hamiltonian of the electron–phonon–electronic system can be written as shown below [1-6]

$$H = H_0 + H_1, \qquad (1)$$

Where

$$H_0 = \sum_{\mathbf{q}} \hbar\omega_{\mathbf{q}} a_{\mathbf{q}}^+ a_{\mathbf{q}} + \sum_{\mathbf{k},\sigma} \varepsilon_{\mathbf{k}} C_{\mathbf{k},\sigma}^+ C_{\mathbf{k},\sigma}, \qquad (2a)$$

$$H_1 = \sum_{\mathbf{k},\mathbf{q},\sigma} (D_{\mathbf{q}} a_{\mathbf{q}} C_{\mathbf{k}+\mathbf{q},\sigma}^+ C_{\mathbf{k},\sigma} + D_{\mathbf{q}}^+ a_{\mathbf{q}}^+ C_{\mathbf{k}-\mathbf{q},\sigma}^+ C_{\mathbf{k},\sigma}), \qquad (2b)$$

Where $\omega_q$ is the frequency of the phonon, $a_q^+$ ($a_q$) is the phonon creation (annihilation) operators, $\varepsilon_k$ is the energy of a electron with a wave vector of $\mathbf{k}$, $C_{\mathbf{k},\sigma}^+(C_{\mathbf{k},\sigma})$ is the electron creation (annihilation) operators, $D_q$ is the coefficient of the electron–phonon–electron interaction.

The Nakajima canonical transformation method can be used to perform the canonical transformation of a multi-body system with the eigenvalue of E ($H\psi = E\psi$) and $H\psi = e^S \psi'$;

hence, the following can be obtained [7]:

$$H_s = e^{-S} H e^S \quad (3).$$

Hamiltonian $H_s$ after transformation satisfies $H_s \psi' = E \psi'$, that is, the eigenvalue is the same.

The quantum mechanics formula is used to record the transformed $H_S$ as follows:

$$\begin{aligned} H_S &= H + [H,S] + [[H,S],S]/2 + \cdots \\ &= H_0 + (H_1 + [H_0,S]) + [H_1,S]/2 + [(H_1 + [H_0,S]),S]/2 + \cdots \end{aligned} \quad (4).$$

A proper S is taken to make $H_1 + [H_0, S] = 0$, eliminate the first order term, maintain the second order term $H_1$ (including all secondary mutual effects of electron–phonon), to eliminate the terms without the phonon operator and electronic self-energy correction terms; and finally, to obtain the effective interaction between electrons, which are given by [1-7]

$$H_{eff} = \sum_{k,k',q,\sigma,\sigma'} V_{kq} C^+_{k+q,\sigma} C_{k,\sigma} C^+_{k'-q,\sigma'} C_{k'\sigma'} \quad (5)$$

Where

$$V_{kq} = \frac{2|D_q|^2 \hbar \omega_q}{(\Delta \varepsilon_{\mathbf{k,q}})^2 - (\hbar \omega_q)^2}$$

The interaction of screening coulomb repulsion between electrons is expressed as[1-7]

$$H_{coul} = \frac{1}{2} \sum_{k_1,k_2,q,\sigma_1,\sigma_2} \frac{4\pi e^2}{q^2 + \lambda^2} C^+_{k_1+q,\sigma_1} C^+_{k_2-q,\sigma_2} C_{k_2,\sigma_2} C_{k_1,\sigma_1}, \quad (6),$$

where $\lambda$ ranges from a few tenths of a nanometer to 1 nanometer.

For the isotropic S-wave scattering, the near attractive effect $\left(\frac{4\pi e^2}{q^2 + \lambda^2} + V_{kq}\right)$ is replaced by the constant $\Xi$, and the total interaction between electrons can be simplified as[1-6]

$$H' \approx -\frac{1}{2} \Xi \sum_{k_1,k_2,q,\sigma_1,\sigma_2} C^+_{k_1+q,\sigma_1} C^+_{k_2-q,\sigma_2} C_{k_2,\sigma_2} C_{k_1,\sigma_1} \quad (7)$$

If $\Xi$ is more than zero, the Hamiltonian can be finally used to obtain the energy gap between the Cooper pair, which is give by[1-6]

$$\Delta = \frac{2\hbar \omega_D}{e^{1/\Xi g(0)} - e^{-1/\Xi g(0)}} \approx 2\hbar \omega_D e^{-\frac{1}{g(0)\Xi}}, g(0)\Xi \ll 1, \quad (8)$$

where $g(0)$ is the density of states near the Fermi surface, $\omega_D$ is the Debye frequency. The energy gap between the Cooper pair is mainly affected by $\omega_D$, $\Xi$, and $g(0)$. If $\Xi$ is larger ($-V_{kq}$ is larger), $\Delta\varepsilon_{\mathbf{k},\mathbf{q}} - \hbar\omega_{\mathbf{q}} \to 0^-$, $V_{kq} \to -\infty$. Thus, this paper studies the equivalent EPEI coefficient $\zeta_{eff} = \sum_{k,k',q,\sigma,\sigma'} V_{kq}$.

## 3. Resonant electron–acoustic (optical) phonon–electron interaction

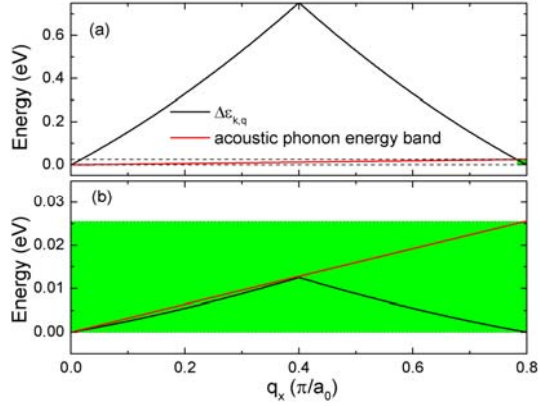

Fig. 2 (a) Non-resonance case; (b) resonance case, the scattering energy difference $\Delta\varepsilon_{\mathbf{k},\mathbf{q}}$ (black solid line) and acoustic phonon energy band (red or gray solid line). In the green region, $\Delta\varepsilon_{\mathbf{k},\mathbf{q}} - \hbar\omega_{\mathbf{q}} < 0$.

We first calculate the two-dimensional electron–acoustic phonon–electron interaction of the common material for comparison. The calculation result is shown in Fig. 2(a). In consideration of the interaction between the electron $k_0 = 0.6\pi/a_0$ in the x direction and other electrons, the electron adopts isotropic free electron approximation $\varepsilon_{\mathbf{k}} = \dfrac{\hbar^2 \mathbf{k}^2}{2m_{eff}}$, where $m_{eff} = m_0$ and $m_0$ refers to the free electron mass. Phonon energy is represented by the approximate linear dispersion $\hbar\omega_q = a_0 \beta_{aph} q / \pi$, where $a_0$ refers to the lattice constant. In the calculation, $\beta_{aph} = 32\text{meV}$ (approximate Si material). In common metal materials, electron effective mass is smaller, and the dispersion of electron energy band is far more than that of the acoustic phonon. $\Delta\varepsilon_{\mathbf{k},\mathbf{q}} - \hbar\omega_{\mathbf{q}} \to 0^-$ can only be realized near the particular intersection point (i.e., within the green region). Electron effective mass $m_{eff}$ can be increased to obtain R-EPEI in a larger scope. The specific calculation result is shown in Fig. 2(b). If $m_{eff} = 60m_0$, $\Delta\varepsilon_{\mathbf{k},\mathbf{q}}$ is slightly smaller than $\hbar\omega_{\mathbf{q}}$ in a larger scope to obtain very strong electron–acoustic phonon–electron interaction.

The proportion of equivalent EPEI coefficients $\zeta_{eff}$ in non-resonance and resonance is about 150:1, that is, R-EPEI is expected to be two multitudes greater than EPEI in non-resonance.

The electron–acoustic phonon interaction is weaker, and the electron energy difference $\Delta\varepsilon_{\mathbf{k,q}}$ in resonance region is smaller. Acoustic phonon energy is always small. The maximum scattering energy difference $\Delta\varepsilon_{\mathbf{k,q}}$ is about half of the maximum acoustic phonon energy. Therefore, the binding energy of electrons caused by the resonance electron–acoustic electron–electron interaction is not too high. Meanwhile, the coulomb interaction between electrons is stronger in the region with a smaller phonon wave vector $\mathbf{q}$ in the resonance area. On the contrary, optical phonon energy is usually higher. The resonance electron–optical phonon–electron interaction will be mainly studied in the following part.

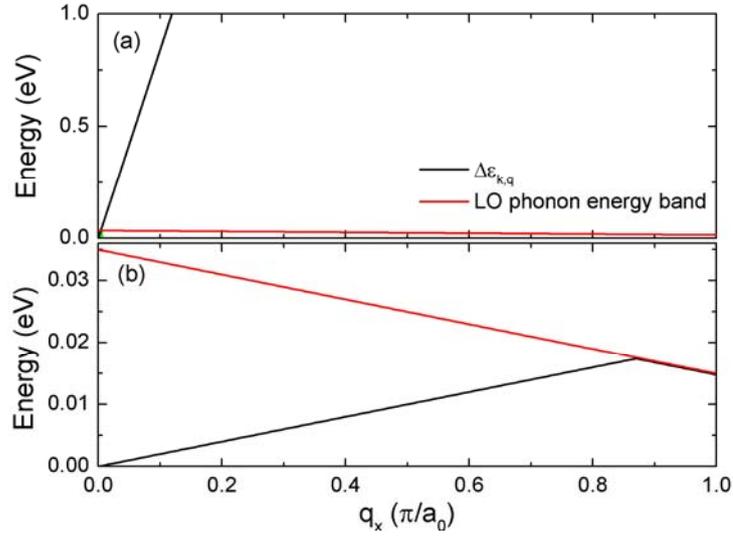

Fig. 3. (a) Non–resonance case; (b) resonance case, scattering energy difference $\Delta\varepsilon_{\mathbf{k,q}}$ (black solid line) and optical phonon energy band (red or gray solid line).

The energy band of the optical phonon is always more complicated than that of the acoustic phonon. For the sake of simplicity, this paper only considers the interaction between one longitudinal optical phonon band and electron. The optical phonon energy is approximately $\hbar\omega_q = \hbar\omega_0 - a_0\beta_{oph}q/\pi$, where $\hbar\omega_0 = 35$ meV and $\beta_{oph} = 20$ meV (similar as the GaAs). In consideration of the interaction between valence (hole) electrons, electron energy is approximately $\varepsilon_{\mathbf{k}} = a_0 v_e q/\pi$. First, non-resonance is considered. Taking $v_e = 8.4$ $eV$ (similar as the graphene or Silicene [13,14]), the calculation result is shown in Fig. 3(a). At the same time, $\Delta\varepsilon_{\mathbf{k,q}} - \hbar\omega_{\mathbf{q}} < 0$, but only in a region with very small q (in the green region). However, the optical phonon-Electron–interaction is stronger than that of the acoustic phonon with this calculation parameter. Therefore, its effective EPEI coefficient $\zeta_{eff}$ will be larger. Resonance will appear in a large scope to decrease $v_e$, e.g., $v_e = 20$ $meV$. The calculation result is shown in Fig. 3(b). When q = 0, optical phonon energy is not equal to 0. When $q \to 0$, $\Delta\varepsilon_{\mathbf{k},\mathbf{q}} \to 0$, resonance will appear in the region with a large q. The proportion of equivalent EPEI coefficient $\zeta_{eff}$ in non-resonance and resonance is about 30:1. Moreover, in non-resonance case, the q in the $\Delta\varepsilon_{\mathbf{k},\mathbf{q}} - \hbar\omega_{\mathbf{q}} < 0$ region is very small, and the coulomb repulsion interaction is stronger. In the resonance case, q is larger, and the coulomb repulsion interaction is weaker. Therefore, R-EPEI is expected to be one multitude greater than the non-resonant EPEI.

## 4. Effect of electron distribution and self-energy correction on R-EPEI

R-EPEI is usually stronger in a particular region of k space because of the characteristics of the electron and phonon energy bands. The result is shown in Fig. 4. Meanwhile, $\Delta\varepsilon_{\mathbf{k},\mathbf{q}} - \hbar\omega_{\mathbf{q}}$ is close to $0^-$ in the x direction. If the electron has a higher electron density of states in the x direction, that is, R-EPEI is stronger when an electron distribution is an anisotropic S-wave or d wave $d_{x^2-y^2}$ distribution (shown in the inset of Fig. 4) [2-4]. The proportion of the equivalent EPEI coefficient $\zeta_{eff}$ of anisotropic distribution to the isotropic distribution is about 2.5:1.

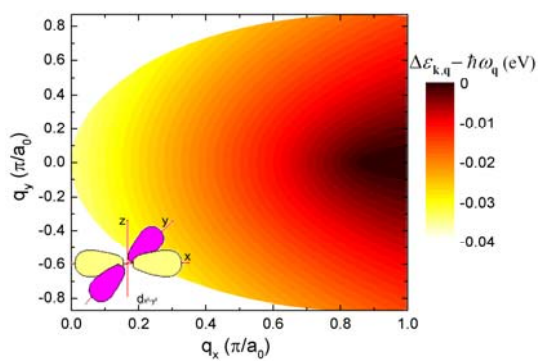

Fig. 4. Change of phonon wave vector $q_x$ with $\Delta\varepsilon_{\mathbf{k},\mathbf{q}} - \hbar\omega_{\mathbf{q}}$. The inset shows the schematic of $d_{x^2-y^2}$ electron wave function.

However, observing R-EPEI in the common material is very difficult. One important consideration is that self-energy correction and EPEI are two-order interactions. In the common material, these two kinds of interactions have the same multitude. Self-energy interaction will cause $|\Delta\varepsilon_{\mathbf{k},\mathbf{q}} - \hbar\omega_{\mathbf{q}}|$ to increase; it will also weaken resonance R-EPEI. Specifically, in consideration of the self-energy process, the electron first emits a phonon with a momentum of $\mathbf{q}$ and then the same phonon is absorbed. If $\varepsilon_{\mathbf{k+q}} - \varepsilon_{\mathbf{k}} < \hbar\omega_{\mathbf{q}}$, the electron self-energy correction can be written as[2-7]:

$$\Delta_{se}\varepsilon_{\mathbf{k+q}} = \sum_{q} \frac{\left|\langle \mathbf{k};1_{\mathbf{q}}|H_1|\mathbf{k+q};0_{\mathbf{q}}\rangle\right|^2}{\varepsilon_{\mathbf{k+q}} - \varepsilon_{\mathbf{k}} - \hbar\omega_{\mathbf{q}}}. \qquad (10).$$

The above formula indicates that this self-energy correction is negative, and that the resonance circumstance appears when $\Delta\varepsilon_{\mathbf{k,q}} - \hbar\omega_{\mathbf{q}} \to 0^{-}$ (in resonance).

The electron self-energy interaction should be weakened to obtain R-EPEI. Particularly, resonant electron self-energy interaction would be eliminated. For example, the anisotropic spin-flip (or intervalley scattering, or pseudospin scattering, etc.) scattering process in k space may weaken the resonance electron self-energy correction. In the EPEI process, a wave vector on the sub-band $B_{\gamma}$ is $\mathbf{k'}$, and spin-down electron $B_{\gamma,\mathbf{k'},\downarrow}$ emits a phonon with a wave vector of $\mathbf{q}$ and transits on sub-band $B_{\chi}$ with a wave vector $\mathbf{k'}-\mathbf{q}$ and spin-upward state $B_{\chi,\mathbf{k'}-\mathbf{q},\uparrow}$. Meanwhile, a spin-upward electron $B_{\alpha,\mathbf{k},\uparrow}$ with a wave vector of $\mathbf{k}$ on sub-band $B_{\alpha}$ absorbs this phonon and transits on sub-band $B_{\beta}$ with a wave vector of $\mathbf{k+q}$ and spin-down state $B_{\beta,\mathbf{k+q},\downarrow}$, as shown in Fig. 1(c). A strong EPEI is expected to appear in resonance. The corresponding self-energy correction is a self-energy process wherein, on the sub-band $B_{\beta}$, the $B_{\beta,\mathbf{k+q},\downarrow}$ electron emits a phonon with a momentum of $\mathbf{q}$, transits on sub-band $B_{\alpha}$ in the state of $B_{\alpha,\mathbf{k},\uparrow}$, and then absorbs the same phonon back to the initial stage. The probability that $B_{\alpha,\mathbf{k},\uparrow}$ electron absorbs phonon to transit on the energy band stage $B_{\beta,\mathbf{k+q},\downarrow}$ is higher, and the probability that $B_{\beta,\mathbf{k+q},\downarrow}$ emits phonon to transit on $B_{\alpha,\mathbf{k},\uparrow}$ is lower or inhibitive, as shown in Fig. 1(d). This means that, in a strong anisotropic spin-flip scattering process in k space, resonance electron self-energy is inhibited and self-energy correction is smaller.

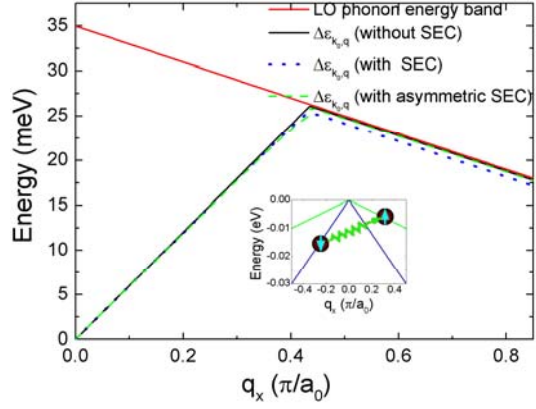

Fig. 5. Effect of electron self-energy correction on $\Delta\varepsilon_{\mathbf{k},\mathbf{q}}$. Self-energy correction is not considered (black solid line); electron self-energy correction is considered (blue dotted line); asymmetric self-energy correction (no spin-flip process self-energy correction, (green dashed line).

The detailed calculation result is shown in Fig. 5. After electron self-energy correction, $\left|\Delta\varepsilon_{\mathbf{k},\mathbf{q}} - \hbar\omega_{\mathbf{q}}\right|$ increases, and the equivalent EPEI coefficient $\zeta_{eff}$ decreases. The proportion of equivalent EPEI coefficient $\zeta_{eff}$ is about 1:0.23:0.73; under three circumstances, namely, no self-energy correction, self-energy correction, and consideration of asymmetric self-energy correction (no spin-flip process self-energy correction).

## 5. Discussion and conclusion

Some characteristics of the superconductor based on R-EPEI are discussed qualitatively.

*Isotope effect.* After the atom is replaced by the isotope, the phonon energy band $\hbar\omega_{\mathbf{q}}$ and $\Delta\varepsilon_{\mathbf{k},\mathbf{q}} - \hbar\omega_{\mathbf{q}}$ is changed. However, after the phonon energy band $\hbar\omega_{\mathbf{q}}$ is changed, R-EPEI does not necessarily disappear. The resonance appears in a different position in the k space, that is, k space would show a very strong isotope effect (it can be positive or negative), but the R-EPEI size would nearly not affected by the isotope effect. This is consistent with the results on high-temperature superconductor [8]. Moreover, in addition to the mass effect, nuclear spin will affect the R-EPEI with spin-flip.

*Phonon density of states.* This is similar as the effect of the electron distribution on the R-EPEI. If the phonon density of stage in resonance region is larger, it is conducive to enhancing R-EPEI. Particularly, R-EPEI will be enhanced greatly if resonance region overlaps with Van Hove singularities position [15-16]. Local phonon may also play an important role in R-EPEI.

*Effect of magnetic field or magnetic interaction.* The magnetic field or magnetic interaction does not necessarily ease R-EPEI. The external magnetic field will change the dispersion relation of the electronic energy band. Furthermore, $\Delta\varepsilon_{\mathbf{k},\mathbf{q}} - \hbar\omega_{\mathbf{q}}$ changes correspondingly. Very strong R-EPEI may appear in the specific magnetic field, generating a peak effect[17-18]. Moreover, the magnetic interaction will enhance the R-EPEI of the spin-flip and inhibit self-energy correction in the spin-flip process.

*Anisotropic S-wave and d wave $d_{x^2-y^2}$.* The electron distribution of anisotropic S-wave and d wave $d_{x^2-y^2}$ can enhance R-EPEI. However, spin-orbit coupling in d electron is stronger than anisotropic S-wave. Furthermore, having spin-flip EPEI is easier and inhibits the self-energy correction in the spin-flip process.

*Experimental verification.* The measurements (calculations) were experimentally performed for the electron and the phonon energy bands in materials (theoretically). The structure of the electron energy band can be regulated and controlled through doping, artificial microstructure (e.g., quantum well), external pressure, external magnetic field, and other methods [2-4,11,19-20]. The phonon energy band can be regulated by using phononic crystals, pressure, and other methods[21-22]. Particularly, the artificial structure of asymmetric phonon in k space has been made conducive to the verification of R-EPEI[23-24]. Therefore, we can determine the proper material and microstructure to achieve R-EPEI experimentally and observe the corresponding superconductive phenomenon.

*The approximate rationality in the model and further improvement.* Numerous approximations were adopted in the calculation to reveal the basic physics in R-EPEI. For example, the electron energy band utilizes free electron or linear dispersion approximation and phonon energy band uses linear dispersion approximation. Some materials have similar electron and phonon energy band structures. However, most materials, particularly high-temperature superconductive materials, have more complicated electron and phonon energy band structures. In addition, the regional phonon mode must be carefully considered after doping. Perturbation approximation is adopted in the calculation of EPEI and self-energy. However, in stronger electron and phonon interactions, perturbation approximation is no longer applicable. EPEI, self-energy, and magnet interaction should be treated united.

In summary, the study shows that EPEI will be increased by 1 to 2 magnitudes in the resonance of electron scattering energy difference and phonon energy. Moreover, anisotropic S-wave electron or $d_{x^2-y^2}$ electron can enhance R-EPEI 2–3 times in resonance. The self-energy correction of electron can weaken R-EPEI, whereas the asymmetric spin-flip scattering process in k space can reduce the effect of electron self-energy to enhance R-EPEI. Qualitatively, the superconductive characteristics caused by R-EPEI show consistency in some aspects with those of the high-temperature superconductor.

Acknowledgements
This work was supported by the NSFC (Grant No. 11364033), the NSF from the Jiangxi Province (Grant No.20122BAB212003), and Science and Technology Project of the Education Department of Jiangxi Province (Grant No. GJJ13005).